\documentclass[twocolumn,showpacs,superscriptaddress,aps,pra]{revtex4-1}
\usepackage{epsfig}
\usepackage{subfigure}
\usepackage{amssymb}
\usepackage{graphicx}
\usepackage{color}
\usepackage{amsfonts}
\usepackage{amscd}
\usepackage{hyperref}
\usepackage{amsmath}
\usepackage{epstopdf}

\usepackage{amsmath}
\usepackage{graphicx}
\usepackage{txfonts}
\usepackage{textcomp}

\usepackage{bm}

\begin{document}

\title{Coherent control techniques in three-level quantum sensing}

\author{Hang Xu}
\affiliation{School of Physics and Optoelectronics Engineering, Anhui University, Hefei 230601, China}

\author{Xue-Ke Song}
\email{songxk@ahu.edu.cn}
\affiliation{School of Physics and Optoelectronics Engineering, Anhui University, Hefei 230601, China}

\author{Dong Wang}
\email{dwang@ahu.edu.cn}
\affiliation{School of Physics and Optoelectronics Engineering, Anhui University, Hefei 230601, China}

\author{Liu Ye}
\affiliation{School of Physics and Optoelectronics Engineering, Anhui University, Hefei 230601, China}

\date{\today }

\begin{abstract}
Quantum coherent control of a quantum system with high-fidelity is rather important in quantum computation and quantum information
processing. There are many control techniques to reach these targets, such as resonant excitation, adiabatic
passages, shortcuts to adiabaticity, and so on. However,
for a single pulse to realize population transfer, the external tiny error has a trivial influence on the final population.
The repeated application of the same pulse will greatly amplify the error effect, making it easy to be detected.
Here, we propose to measure small control errors in three-level quantum systems by coherent amplification of their effects, using
several coherent control techniques.
For the two types of Hamiltonian with SU(2) dynamic symmetry, we analyze how the fidelity of population transfer are affected by Rabi frequencies fluctuation and static detuning deviation,
based on the pulse sequence with alternating and same phases, respectively. It is found that the sensitivity of detecting these errors can be effectively amplified by the control pulse sequences.
Furthermore, we discuss the efficiency of sensing the two errors with these control techniques by comparing the full width at half maximum of the population profiles.
The results provide an accurate and reliable way for sensing the weak error in three-level quantum systems by applying repeatedly the coherent control pulse.

\end{abstract}

\maketitle

\section{Introduction}
\label{sec0}


Quantum coherent manipulation of a quantum state is a fundamental prerequisite in atomic and molecular physics \cite{atom1}, optics \cite{optics1,optics2}, chemistry \cite{chemistry1}, quantum information and computation\cite{information1,information2}. In particular, it has many practical applications in the field of quantum information science in recent years, including preparation of quantum gates and quantum states \cite{gate1}, molecular chirality resolution \cite{chiral1}, quantum batteries \cite{battery1}, detection of parity violation \cite{parity1}, and so on. Also, the research related to this topic has also been expanded from closed systems to open systems \cite{open1}. To reach universal quantum computation,
the prepared quantum gate or quantum state is required to have a ultra-high fidelity (typically with an error in the range of ${10^{ - 2}}$-${10^{ - 4}}$) under the influence of parameter fluctuations of the systems or external environmental noises.

Lots of protocols have been proposed to achieve an accurate and robust quantum coherent control with ultra-high fidelity, such as resonant pulses \cite{RE}, adiabatic passages \cite{RAP1,RAP2,PAP1,STIRAP1,STIRAP2}, shortcuts to adiabaticity (STA) \cite{LRI0,LRI1,song1,song2,FF1,LRI2,LRI3,LRI4,LRI5}, optimal control theory \cite{OCT1,OCT2,OCT3,OCT4}, dressed states driving \cite{DSD1,DSD2,DSD3}, composite pulses \cite{CAP1,UCP1,CP1}, and so on. Recently, these protocols have attracted much attention in both theory and experiment
\cite{E-CP,E-Rydberg,E-NV,E-OCT,E-IGO,E-open}.
In 2015, Barredo \emph{et al.} \cite{E-Rydberg} investigated coherent excitation hopping in a spin chain of three Rydberg atoms in experiment.
In 2015, N\"{o}bauer \emph{et al.} \cite{E-NV} experimentally achieved tailored robustness against qubit inhomogeneities and control errors using quantum optimal control technique.
In 2017, Damme \emph{et al.} \cite{E-OCT} established time- and energy-minimum optimal control strategies for the robust and precise state control of two-level quantum systems.
In 2021, Wu \emph{et al.} \cite{E-open} proposed a fast mixed-state control scheme to transfer the quantum state along designable trajectories in Hilbert space by STA.

In fact, the preparation of quantum states and quantum gates is usually sensitive to external environmental disturbances
and parameter fluctuations of Hamiltonian \cite{error1,error2}. However, some control errors originating from specific
experimental implementations and conditions cannot be avoided in experiment. Therefore, it is fairly necessary to
find a simple and reliable method to characterize such unwanted experimental errors. Recently, some protocols
measuring small parameter errors in quantum systems by coherent amplification of their effect for
quantum sensing of weak electric and magnetic fields \cite{E-sense1,sense1,E-sense2,E-sense3,E-sense4}.
For instance, in 2015, Ivanov \emph{et al.} \cite{E-sense1} introduced a quantum-sensing protocol for detecting very small forces assisted by a symmetry-breaking adiabatic transition in the quantum Rabi model.
In 2018, Ivanov \emph{et al.} \cite{E-sense2} proposed a quantum sensing scheme for measuring the phase-space-displacement parameters using a single trapped ion.
In 2021, Zhang \emph{et al.} \cite{E-sense3} presented a quantum sensing model using color detuning dynamics with dressed states driving based on the stimulated Raman adiabatic passage.
In 2021, Vitanov \cite{E-sense4} shown a quantum sensing of weak electric and magnetic fields by coherent amplification of the effects of small energy level shifts.

In this paper, we propose a quantum sensing method to detect the error effects originating from parameter fluctuations of Hamiltonian in three-level quantum systems, using some coherent control pulses. The error effects are mapped on
to change speed of the populations of the quantum states by repeatedly applying quantum gate operations. In principle, the sensitivity of sensing the errors of a single pulse is usually very small, the error effect can be coherent amplified by applied repeatedly the same pulse in a specific composite pulse sequence. For Rabi frequencies fluctuation and static detuning deviation, we employ the pulse sequence with alternating and same phases to study the sensitivity of a population transfer, respectively. The results show that the sensitivity of sensing the two errors can be greatly improved. Furthermore, we
give a detailed comparison about the error effects in different coherent control techniques, and accordingly find that there exists a different sequential order in these control techniques in terms of different errors.

The paper is organized as follows. In Sec. \ref{sec2}, we briefly introduce the three-level Hamiltonian model and the two different pulse sequences. In Sec. \ref{sec3}, we analyze the three-level Hamiltonian in the case of off-resonance and use the two pulse sequences to detect two kinds of errors: Rabi frequency error and static detuning deviation. In Sec. \ref{sec4}, the error effect from Rabi frequency error in the one-photon resonance case is studied. We finally show the discussion and summary in Sec. \ref{sec5}.

\section{Preliminaries}
\label{sec2}
\begin{figure}[tbp]
  \centering
  \includegraphics[width=1\linewidth]{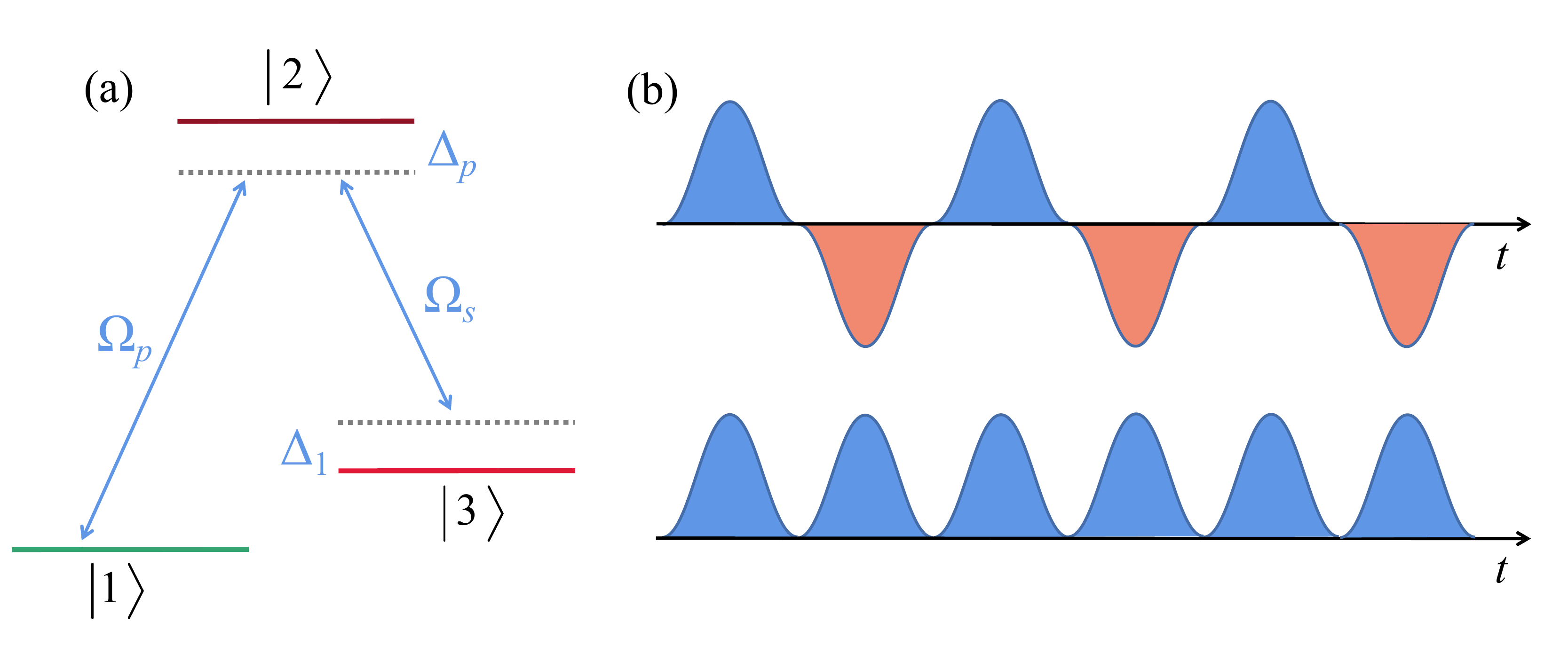}
  \caption{(a) $\Lambda$-type three-level system, where ${\Omega _p}$ and ${\Omega _s}$ are Rabi frequencies of pump and Stokes pulses, and ${\Delta _p}$ and ${\Delta _1}$ are the detunings from the resonances.
  (b) Schematic diagram for the pulse sequence with alternating  phases (top) and same phases (bottom), respectively. In this case, the pump and Stokes pulses take the same pulse profiles.}
  \label{Fig1}
\end{figure}

A general $\Lambda$-type three-level system driven by external coherent fields can be described by the Schr\"{o}dinger equation,
\begin{eqnarray}
i\hbar {\partial _t}\bm{\psi(t)} = {H_0}(t)\bm{\psi(t)},
\end{eqnarray}
where $\bm{\psi(t)} = {[{c_1}(t),{c_2}(t),{c_3}(t)]^T}$ is a column vector with the probability amplitudes of the three bare states $\left| 1 \right\rangle $, $\left| 2 \right\rangle $, and $\left| 3 \right\rangle $.
The Hamiltonian, under the rotating wave approximation, on the basis of the bare states  $\left\{ {\left| 1 \right\rangle ,\left| 2 \right\rangle ,\left| 3 \right\rangle } \right\}$ is
\begin{eqnarray}
{H_0} = \frac{\hbar }{2}\left( {\begin{array}{*{20}{c}}
0&{{\Omega _p}{e^{ - i\alpha }}}&0\\
{{\Omega _p}{e^{i\alpha }}}&{2{\Delta _p}}&{{\Omega _s}{e^{ - i\beta }}}\\
0&{{\Omega _s}{e^{i\beta }}}&{2{\Delta _1}}
\end{array}} \right),\label{H0}
\end{eqnarray}
where ${\Omega _p}$ and ${\Omega _s}$ are Rabi frequencies of pump ($P$) and Stokes ($S$) pulses or electromagnetic fields, and $\alpha $ and $\beta $ are phases of the pulses, respectively, shown in Fig. \ref{Fig1} (a). The detunings, respectively, are defined as ${\Delta _p} = ({E_2} - {E_1})/\hbar  - {\omega _p}$, ${\Delta _s} = ({E_2} - {E_3})/\hbar  - {\omega _s}$, and ${\Delta _1} = {\Delta _p} - {\Delta _s}$, where ${\omega _p}$ and ${\omega _s}$ are drive frequencies of pulses. ${E_n}$ ($n = 1,2,3$) are the energies of bare state. Levels $\left| 1 \right\rangle$ and $\left| 2 \right\rangle$ are coupled by $P$ pulse, and levels $\left| 2 \right\rangle$ and $\left| 3 \right\rangle$ are coupled by $S$ pulse. On two photon resonance ($\Delta _1=0$), the technique of the stimulated Raman adiabatic passage can be used to realize a high-fidelity population transfer from the state $\left| 1 \right\rangle $ to the state $\left| 3 \right\rangle $ along the adiabatic dark state, based on the counterintuitive order of pulses. We assume $\alpha  = \beta $ in the context. By tuning the control parameters of the Hamiltonian in Eq. (\ref{H0}), we can obtain two families of Hamiltonian with the SU(2) dynamic symmetry: CASE (I) Off-resonance Hamiltonian with Majorana decomposition when $\Omega={\Omega _p} = {\Omega _s}$ and $\Delta={\Delta _p} =  - {\Delta _s}$, and CASE (II) One-photon resonance Hamiltonian when ${\Delta _p} = {\Delta _1} = 0$. In principle, both of them can be reduced to the effective two-level Hamiltonian, while they show different performances on the control of quantum states.
In the following, we shall discuss these two cases separately.

In the ideal case, many methods with different pulses, including adiabatic and nonadiabatic ones, can be used to drive the evolution of a quantum system to achieve a complete population transfer. But if there exists a weak error or perturbation on the control of the system, such as Rabi frequency and detuning errors, the efficiency and fidelity of the transfer will be reduced. Therefore, we can predict the existence of the error by measuring the populations of states. However, for a single pulse, the experimental error usually has a small effect on the populations, it is effective to use composite pulse schemes to amplify the error effects.

To sense the experimental errors originating from specific experimental implementations and conditions: Rabi frequencies fluctuation and static detuning deviation with the SU(2) dynamic symmetry Hamiltonian, we examine how the fidelity of population transfer is affected by these errors, based on two simple composite pulse sequences: the alternating and same phases of pulse sequence. The alternating phases of pulse sequence is that the same pulse is again applied $N$ times but flips the phase by $\pi$ from pulse to pulse [see Fig. \ref{Fig1} (b) top]. That is, a minus sign is applied to the two Rabi frequencies of every even-numbered pulse in the sequence simultaneously. The same phases of pulse sequence is defined as that the same pulse is applied $N$ times, where $N$ is used to represent the total number of pulses [see Fig. \ref{Fig1} (b) bottom].

\section{CASE (i): Off-resonance}
\label{sec3}
Here, we study the case that $\Omega={\Omega _p} = {\Omega _s}$ and $\Delta={\Delta _p} =  - {\Delta _s}$, the Hamiltonian ${H_0}$ is reduced to the Hamiltonian ${H_1}$, which possesses the SU(2) dynamic symmetry. In this case, the Hamiltonian of the system can be expressed as
\begin{eqnarray}
{H_1} = \frac{\hbar }{2}\left( {\begin{array}{*{20}{c}}
{ - 2\Delta }&{\Omega {e^{ - i\alpha }}}&0\\
{\Omega {e^{i\alpha }}}&0&{\Omega {e^{ - i\alpha }}}\\
0&{\Omega {e^{i\alpha }}}&2\Delta
\end{array}} \right),
\end{eqnarray}
which can also be constructed by the two Heisenberg-interacting spins model \cite{LRI2} and trapped ion system \cite{H2}.
This three-level system can be reduced to a two-level system with the ground state $\left| g \right\rangle $ and excited state $\left| e \right\rangle $ by Majorana decomposition \cite{H1}, which gives
\begin{eqnarray}
H = \frac{\hbar }{2}\left( {\begin{array}{*{20}{c}}
{ - \Delta }&{\Omega {e^{ - i\alpha }}/\sqrt 2 }\\
{\Omega {e^{i\alpha }}/\sqrt 2 }&{  \Delta }
\end{array}} \right).\label{efftwo}
\end{eqnarray}

For the Hamiltonian $H$, its propagator of a pulse with zero phase describing the evolution
of the corresponding two-state system is expressed, in terms of the Cayley-Klein parameters $a$ and $b$, as
\begin{eqnarray}
U= \left( {\begin{array}{*{20}{c}}
a&b\\
{ - {b^*}}&{{a^*}}
\end{array}} \right).\label{real}
\end{eqnarray}
By mapping it onto the three state system, the propagator for the ${H_1}$ is
\begin{eqnarray}
{U_1}= \left( {\begin{array}{*{20}{c}}
{{a^2}}&{\sqrt 2 ab}&{{b^2}}\\
{ - \sqrt 2 a{b^*}}&{{{\left| a \right|}^2} - {{\left| b \right|}^2}}&{\sqrt 2 {a^*}b}\\
{ - {b^{*2}}}&{ - \sqrt 2 {a^*}{b^*}}&{{a^{*2}}}
\end{array}} \right).
\end{eqnarray}
We can find that the populations of the energy levels $\left| 1 \right\rangle $ and $\left| 3 \right\rangle $ in the three-level system are similar to those of $\left| g \right\rangle $ and $\left| e \right\rangle $  in the two-level system, i.e.,
\begin{eqnarray}
\begin{array}{l}\vspace{0.06cm}
{P_1} = {\left| a \right|^4} = p^2(g),\\
{P_3} = {\left| b \right|^4} = p^2(e),
\end{array}
\end{eqnarray}
where ${P_n}$ $(n=1,2,3)$ are level populations of $\left| n \right\rangle $ in three-level systems with ${H_1}$, and ${p_k}$ $(k=g,e)$ are level populations of $\left| k \right\rangle $ in two-level systems with $H$. Thus the population transfer of $\left| 1 \right\rangle  \to \left| 3 \right\rangle $ driven by ${H_1}$ can be treated as the population inversion of $\left| g \right\rangle  \to \left| e \right\rangle $ driven by $H$. In the following, we study the quantum sensing of the change of the electric and magnetic field by amplifying the population shifts in two different ways: Rabi frequency error and static detuning deviation.

\subsection{Rabi frequency error}

 When the Rabi frequency errors occur on both $P$ pulse and $S$ pulse in this three-level system, this corresponds to that the Rabi frequency $\Omega$ in ${H_1}$ has a error. We use a small time-independent and dimensionless parameter $\sigma $ to represent error-inducing Rabi frequency variation, i.e., $\Omega  \to (1 + \lambda )\Omega $. To begin with, let us analyze the feasibility of alternating phases pulse sequence. For the Hamiltonian $H$ in two-level systems, the propagator of a pulse with $\pi$ phase is
\begin{eqnarray}
{U_ - } = \left( {\begin{array}{*{20}{c}}
a&{ - b}\\
{{b^*}}&{{a^*}}
\end{array}} \right).
\end{eqnarray}
When the parameters of the Hamiltonian satisfy the following symmetry,
\begin{eqnarray}
\Delta (t) =  - \Delta ({t_f} - t),\;\;\;\Omega (t) = \Omega ({t_f} - t).\label{symmetry}
\end{eqnarray}
the parameter of the propagator of a single pulse $a$  is a real number \cite{a}. As a result, we have
\begin{eqnarray}
U = \left( {\begin{array}{*{20}{c}}
a&b\\
{ - {b^*}}&a
\end{array}} \right),\;\;\;{U_ - } = \left( {\begin{array}{*{20}{c}}
a&{ - b}\\
{{b^*}}&a
\end{array}} \right).
\end{eqnarray}
We can find that $U{U_ - } = {U_ - }U=I$, implying that the whole pulse sequence is acting as an identity matrix if we use the even number of the pulses in the sequence. The odd number of the pulses in the sequence is acting as a single pulse. Thus, it is hard to use a pulse sequence with alternating phases to improve the sensitivity to Rabi frequency error.

Now, let us consider the pulse sequence with same phases by using four different pulse techniques.
Firstly, we use the simplest model, $\pi$ pulses, to sense weak electromagnetic fields by amplifying the effects of Rabi frequency error.
The parameters of Hamiltonian in Eq. (\ref{efftwo}) can be chosen as
\begin{eqnarray}
\Delta (t) = 0,\;\;\;\int_{{t_0}}^{{t_f}} {\Omega (t)/\sqrt 2 dt = \pi } ,
\end{eqnarray}
and the propagator of a single pulse can be expressed analytically as
\begin{eqnarray}
U = \left( {\begin{array}{*{20}{c}}
{ - \sin \frac{{\lambda \pi }}{2}}&{ - i\cos \frac{{\lambda \pi }}{2}}\\
{ - i\cos \frac{{\lambda \pi }}{2}}&{ - \sin \frac{{\lambda \pi }}{2}}
\end{array}} \right).
\end{eqnarray}
The total propagator of any number of pulses $N$ can be expressed as
\begin{eqnarray}
{U^{2n}} = \left( {\begin{array}{*{20}{c}}
{{{( - 1)}^n}\cos n\lambda \pi }&{i{{( - 1)}^{n + 1}}\sin n\lambda \pi }\\
{i{{( - 1)}^{n + 1}}\sin n\lambda \pi }&{{{( - 1)}^n}\cos n\lambda \pi }
\end{array}} \right)
\end{eqnarray}
and
\begin{eqnarray}
{U^{2n + 1}} = \left(\! {\begin{array}{*{20}{c}}
{{{( - 1)}^{n + 1}}\sin \frac{{2n + 1}}{2}\lambda \pi }&{i{{( - 1)}^{n + 1}}\cos \frac{{2n + 1}}{2}\lambda \pi }\\
{i{{( - 1)}^{n + 1}}\cos \frac{{2n + 1}}{2}\lambda \pi }&{{{( - 1)}^{n + 1}}\sin \frac{{2n + 1}}{2}\lambda \pi }
\end{array}}\! \right),
\end{eqnarray}
where $N=2n$ and $N=2n+1$ donate the number of pulses is even and odd, respectively. Therefore, for the three-level system with Hamiltonian ${H_1}$, the population transfer from
$\left| 1 \right\rangle $ to $\left| 3 \right\rangle $
can be realized by applying odd pulses, and it gives
\begin{eqnarray}
{P_3} = {\cos ^4}(N\lambda \pi/2).
\end{eqnarray}
When even pulses are applied, the population will remain in energy level $\left| 1 \right\rangle $ with
\begin{eqnarray}
{P_1} = {\cos ^4}(N\lambda \pi/2).
\end{eqnarray}
\begin{figure}[tbp]
  \centering
  \includegraphics[width=1.08\linewidth]{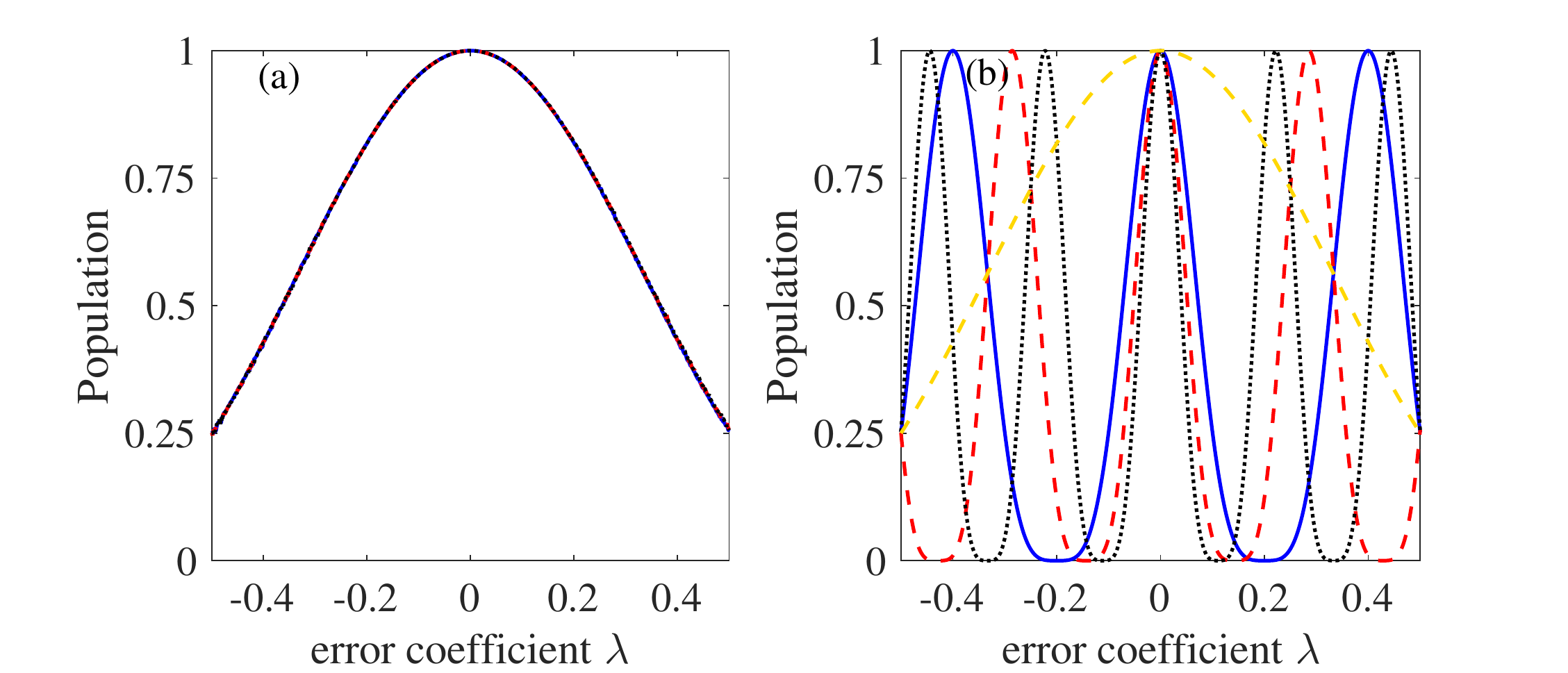}
  \caption{Population of level $\left| 3 \right\rangle $ vs the Rabi frequency error coefficient $\lambda $ for sequences of $N=5,$ $7,$ and $9$ identical pulses, where $N=5$ (blue, solid line), $N=7$ (red, dashed line), and $N=9$ (black, dotted line). (a) $\pi$ pulse sequence with alternating phases. (b) $\pi$ pulse sequence with same phases, together with a single $\pi$ pulse (yellow, dashed-dotted line).}
  \label{Fig2}
\end{figure}
One can see that with the increase of the number of pulses $N$, the evolve period of the population $P(1)$ and $P(3)$ decrease, and the change speed of the population increases near $\lambda  = 0$, that is, the sensitivity with respect to Rabi frequency increases.
In Fig. \ref{Fig2}, we plot the population $P(3)$ with the change of the Rabi frequency error for $N=5$, $7$, and $9$, based on $\pi$ pulse sequence with alternating and same phases, respectively. It can be seen that the pulse sequence with alternating phases cannot improve the sensitivity of sensing the Rabi frequency error. For the single pulse, the profile of population is much broader than the sequences of pulses of same phases with $N=5$, $7$, and $9$ near $\lambda =0$. Moreover, the narrow features produced by the pulse sequence with same phases becomes more narrow as the number of pulses increases.

\begin{figure}[tbp]
  \centering
  \includegraphics[width=1.08\linewidth]{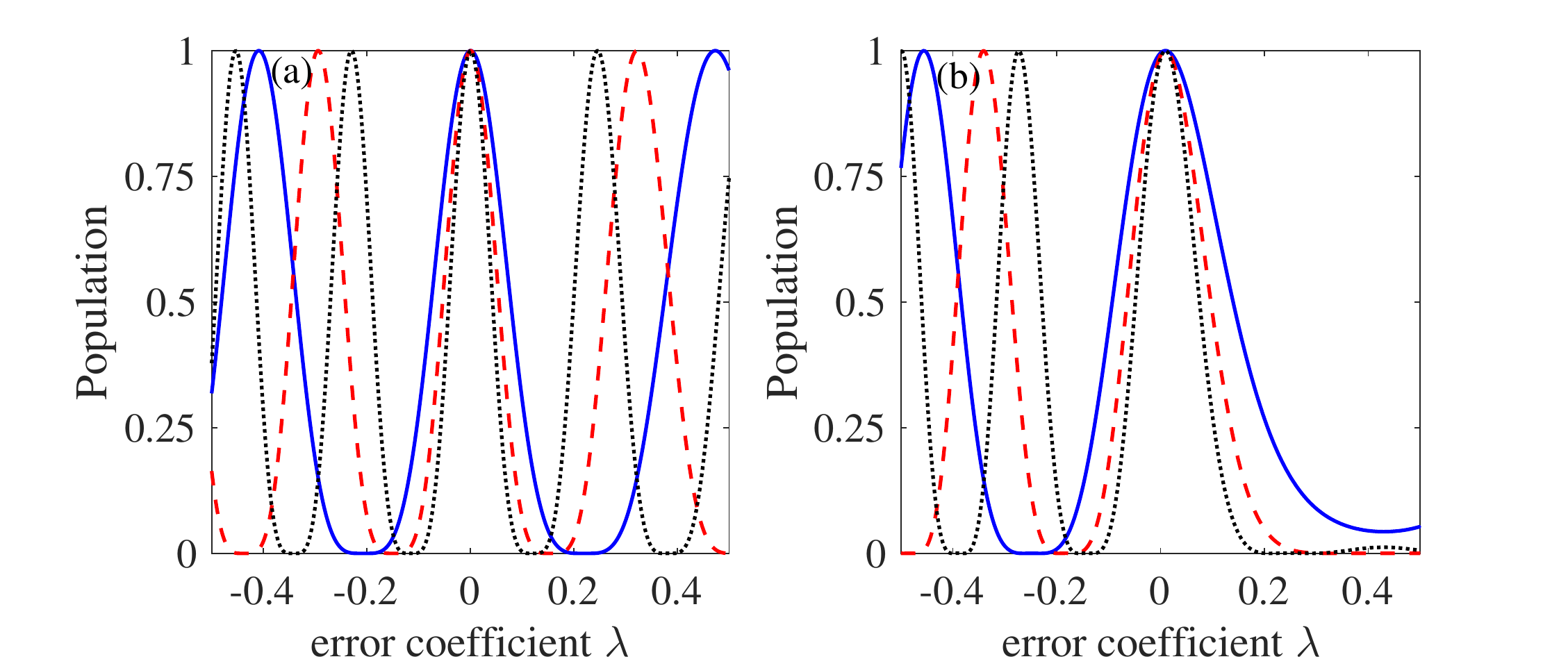}
  \caption{Population of level $\left| 3 \right\rangle $ vs the Rabi frequency error coefficient $\lambda $ for sequences of $N=5,$ $7,$ and $9$ identical pulses, where $N=5$ (blue, solid line), $N=7$ (red, dashed line), and $N=9$ (black, dotted line). (a) The pulse sequence with same phases using Gaussian pulses. (b) The pulse sequence with same phases using AE adiabatic passages.}
  \label{Fig3}
\end{figure}

Next, we use Gaussian pulses with linear chirp \cite{Gaussian} and Allen-Eberly (AE) adiabatic passages which is a special case of the level-crossing Demkov-Kunike model \cite{AE} to study how the population is affected by the Rabi frequency error. For Gaussian pulses with linear chirp, the parameters of the Hamiltonian are
\begin{eqnarray}
\Delta (t) = \sqrt 2 t/{T^2}, \;\;\;\;\Omega (t) = 2\sqrt 2{e^{ - {{(t/T)}^2}}}/T,
\end{eqnarray}
where T is the pulse width. For the AE adiabatic passages, the Hamiltonian parameters are set as
\begin{eqnarray}
\Delta (t) = \sqrt 2  \tanh (t/T) /T,\;\;\;\;\Omega (t) = \sqrt 6  {\mathop{\rm sech}\nolimits} (t/T) /T.
\end{eqnarray}
In Fig. \ref{Fig3}, we plot the population $P(3)$ with the change of the Rabi frequency error for $N=5$, $7$, and $9$, based on Gaussian pulse and AE adiabatic pulse sequence with same phases, respectively. It can be observed that the sensitivity of sensing the Rabi frequency error increases as the number of pulses increases. Furthermore, at around of $\lambda   = 0$, the full width of half maximum of
the population for AE adiabatic passages is larger than the one for Gaussian pulses, suggesting that Gaussian pulses are more sensitive to sense the Rabi frequency error than AE adiabatic passages.
In fact, the adiabaticity of AE adiabatic passages is better than Gaussian pulses, thus AE adiabatic passages is more stable than the latter.

\begin{figure}[tbp]
  \centering
  \includegraphics[width=0.75\linewidth]{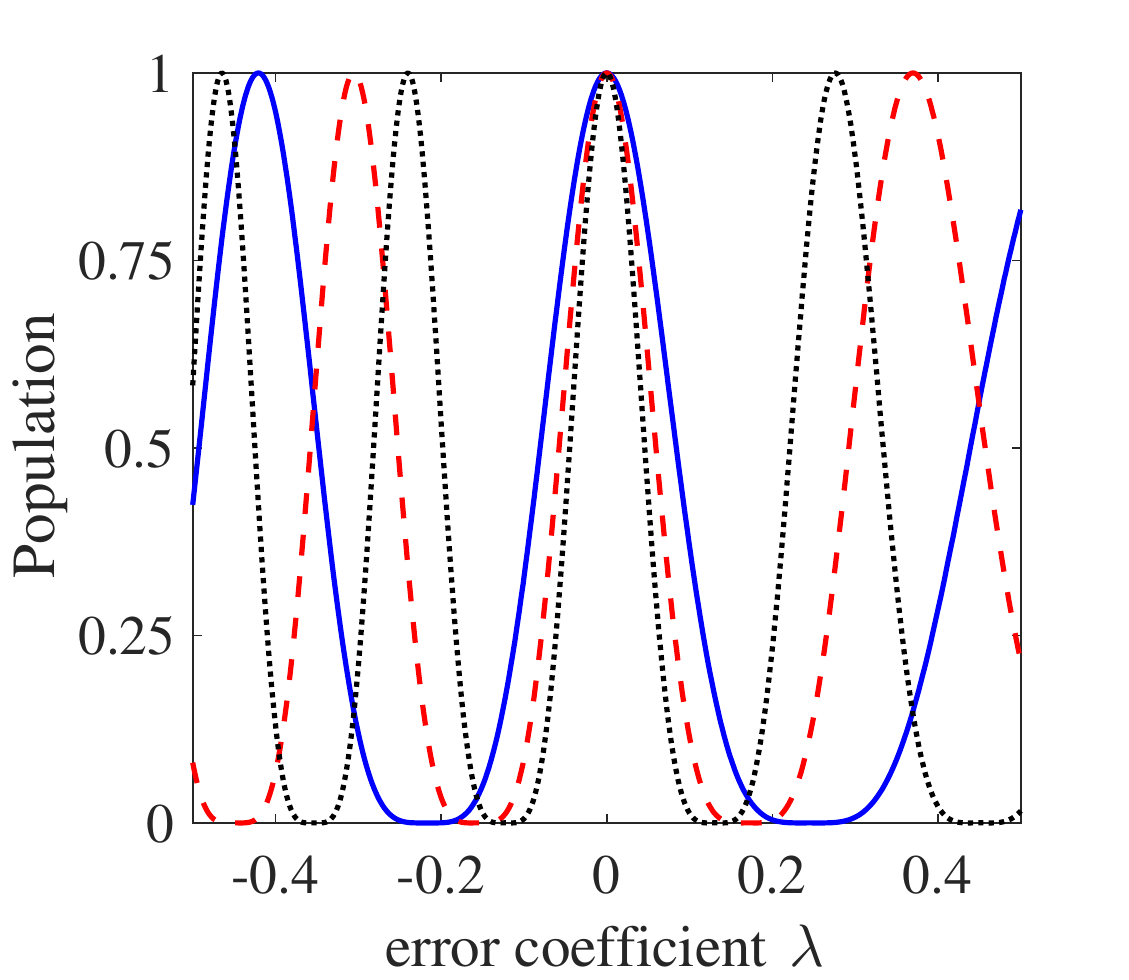}
  \caption{Population of level $\left| 3 \right\rangle $ vs the Rabi frequency error coefficient $\lambda $ for pulse sequences with same phases of $N=5,$ $7,$ and $9$ identical pulses using STA, where $N=5$ (blue, solid line), $N=7$ (red, dashed line), and $N=9$ (black, dotted line).}
  \label{Fig4}
\end{figure}

Finally, we use STA technique to analyze the sensitivity of a population transfer via Rabi frequency error. For the Hamiltonian $H(t)$, there is an invariant \cite{LRI0}
\begin{eqnarray}
I = \left( {\begin{array}{*{20}{c}}
{\cos \theta }&{\sin \theta {e^{ - i\gamma  }}}\\
{\sin \theta {e^{i\gamma }}}&{ - \cos \theta }
\end{array}} \right),
\end{eqnarray}
which satisfies the dynamical equation
\begin{eqnarray}
\frac{{dI}}{{dt}} = \frac{{\partial I}}{{\partial t}} + \frac{1}{{i\hbar }}\left[ {I,H} \right] = 0.
\end{eqnarray}
By solving this equation, we obtain the constraint conditions
\begin{eqnarray}
\dot \theta  =  - \Omega \sin \gamma /\sqrt 2,\;\;\;{\rm{     }}\dot \gamma {\rm{ = }} - \Omega \cot\theta \cos\gamma /\sqrt 2 - \Delta .\label{rabiyuanshi}
\end{eqnarray}
Using the eigenstates ${\left| {{\varphi _n}(t)} \right\rangle }$ of the invariant $I$, the solution of Schr\"{o}dinger equation $i\hbar\partial\left| {\psi (t)} \right\rangle/\partial t=H(t)\left| {\psi (t)} \right\rangle$ \cite{LRI0} can be written as
\begin{eqnarray}
\left| {\psi (t)} \right\rangle  = \sum\limits_ {j=\pm}  {{C_j}{e^{i{\epsilon _j}(t)}}\left| {{\varphi _j}(t)} \right\rangle } ,
\end{eqnarray}
where ${{C_j}}$ are constants, and ${\epsilon_j}(t)$ are the LR phases with

\begin{eqnarray}
{{\dot \epsilon }_ + }(t) =  - {{\dot \epsilon }_ - }(t) = \frac{{\dot \theta \cos\gamma }}{{2\sin \theta \sin \gamma }}.\label{neta}
\end{eqnarray}
In order to realize population inversion along the eigenstate $\left| {{\varphi _ + }(t)} \right\rangle$, one can choose the $\theta (t)$ as
\begin{eqnarray}
\theta (t) = \pi t/T.
\end{eqnarray}
Here, we try the Fourier series type of Ansatz
\begin{eqnarray}
{\epsilon _ + }(t) =  - \theta  - n\sin (2\theta ),
\end{eqnarray}
where $n$ is a freely chosen parameter. Using the Eq.~(\ref{neta}), we obtain
\begin{eqnarray}
\gamma (t) =  - {\rm{arccot}}[2(1 + 2n \cdot \cos2\theta )\sin\theta ].
\end{eqnarray}
Based on the perturbation expansion \cite{LRI3}, we can define the error sensitivity ${q_s}$ with respect to the Rabi frequency error $\lambda$ as
\begin{eqnarray}
{q_s} =  - \frac{{{\partial ^2}P({t_f})}}{2{\partial {\lambda ^2}}}{|_{\lambda  = 0}},
\end{eqnarray}
where $P({t_f})$ is the transition probability at the final time. The ${q_s}$ is determined by selecting $n$.  The larger the value of ${q_s}$, the more sensitive it is.
Here, we choose $n=0.5$, and this gives a relatively large value of ${q_s}$ with 1.91.
In Fig. \ref{Fig4}, we plot $P(3)$ with the change of the Rabi frequency error for $N=5$, $7$, and $9$, based on STA pulse sequence with same phases. The number of pulses in STA pulses sequence
can affect the efficiency of population transfer, and the sensitivity increases with the increase of the number of pulses.

Finally, we discuss effect of sensing the Rabi frequency error in these different schemes with same phases by comparing the full width at half maximum of the population change curve near $\lambda  = 0$, based on the same numbers of pulses. This is shown in Table \ref{Table1}. The smaller the value of the full width at half maximum of population, the higher the sensitivity of the scheme with Rabi frequency error.
For the same numbers of pulses, the sensitivity to Rabi frequency error of these schemes shows a sequential order: $\pi$ pulse, Gaussian pulse, STA pulse, and AE adiabatic pulse sequences.

\begin{table}[htb]
\centering \caption{ Full width at half maximum of population in different schemes to sense the Rabi frequency error. We also use -S to indicate the the pulse sequence with same phases.}
\begin{tabular}{ccccc}
\hline\hline
Number of pulses     \;\;\;\; &  $\pi$-S      \;\;\;\;\;&  Gaussian-S        \;\; \;\; &  AE-S     \;\;\;\;\;    &  STA-S   \\
   \hline
$N=5$      \;\;\;\;\; &  0.146            \;\;\;\;\; & 0.153          \;\;\;\;\;  &  0.233        \;\;\;\;\; &  0.165         \\
$N=7$      \;\;\;\;\; &  0.104            \;\;\;\;\;& 0.109          \;\;\;\;\;  &  0.162        \;\;\;\;\; &  0.117        \\
$N=9$      \;\;\;\;\; &  0.082            \;\;\;\;\;& 0.084          \;\;\;\;\;  &  0.123        \;\;\;\;\; &  0.091        \\
\hline\hline
\end{tabular}\label{Table1}
\end{table}

\subsection{Static detuning deviation}


Now let us consider the case that there exists a small static detuning deviation on the Hamiltonian $H_1$, i.e., $\Delta  \to \Delta  + \delta $, where $\delta $ is static detuning. This may be caused by fluctuating frequency of the laser, microwave, or radio-frequency generator.
Due to the existence of static detuning, the Hamiltonian parameters do not satisfy symmetry in Eq. (\ref{symmetry}), and the Cayley-Klein parameter $a$ is no longer a real number. Its propagator
is described by the most general form in Eq. (\ref{real}).

For the pulse sequence with alternating phases, the propagator of two-level system after using two pulses is
\begin{eqnarray}
U_ \pm ^2 = {U_ - }U = \left( {\begin{array}{*{20}{c}}
{{a^2} + {{\left| b \right|}^2}}&{2ib{a_i}}\\
{2i{b^*}a}&{{a^{*2}} + {{\left| b \right|}^2}}
\end{array}} \right).
\end{eqnarray}
Thus we can write the propagator of any even number of pulses $N=2n$ with alternating phases in terms of
\begin{eqnarray}
\begin{array}{l}
U_ \pm ^{2n} = {(U_ \pm ^2)^n}\\\;\;\;\;\;\;
 = \left( {\begin{array}{*{20}{c}}
{\cos n\Theta  + 2i{a_r}{a_i}E}&{2ib{a_i}E}\\
{2i{b^*}{a_i}E}&{\cos n\Theta  - 2i{a_r}{a_i}E}
\end{array}} \right),
\end{array}
\end{eqnarray}
where $\Theta  = \arccos (1 - 2a_i^2)$ and $E = \sin n\Theta/\sin \Theta$.
In addition, the propagator of any odd number of pulses $N=2n+1$ with alternating phases can be expressed as
\begin{eqnarray}
U_ \pm ^{2n + 1} = UU_ \pm ^{2n} = \left( {\begin{array}{*{20}{c}}
{aF + 2i{a_i}E}&{bF}\\
{ - {b^*}F}&{aF - 2i{a_i}E}
\end{array}} \right),
\end{eqnarray}
where $F = \cos (n + 1/2)\Theta/\cos (\Theta /2)$. Therefore, the populations of $\left| 1 \right\rangle $ when $N$ is even and $\left| 3 \right\rangle $ when $N$ is odd are, respectively,
\begin{eqnarray}
\begin{array}{l}\vspace{0.1cm}
{P_1}= {\left| b \right|^4}{\cos ^4}(N\Theta /2)/{\cos ^4}(\Theta /2),\;\;\;N = even;\\
{P_3}= {\left| b \right|^4}{\cos ^4}(N\Theta /2)/{\cos ^4}(\Theta /2),\;\;\;N = odd.
\end{array}\label{alternating}
\end{eqnarray}

On the other hand, when we consider the pulse sequence with same phases in this disturbed two-level system, the total propagator is the multiplication of $N$ individual propagators $U$, that is
\begin{eqnarray}
{U^N} = \left( {\begin{array}{*{20}{c}}
{\cos N\vartheta  + i{a_r}D}&{bD}\\
{ - {b^*}D}&{\cos N\vartheta  - i{a_r}D}
\end{array}} \right),
\end{eqnarray}
where ${a_r}$ and ${a_i}$ are the real and imaginary parts of $a$, $\vartheta  = \arccos ({a_r})$, and $D =\sin N\vartheta/\sin \vartheta$. The population of the excited state $\left| e \right\rangle $
in two-level system after using  pulse sequence with same phases is
\begin{eqnarray}
{p_e} = {\left| b \right|^2}\frac{{{{\sin }^2}N\vartheta }}{{{{\sin }^2}\vartheta }}.
\end{eqnarray}
Therefore, in the three-level system, the corresponding population of the states $\left| 3 \right\rangle $ and $\left| 1\right\rangle $, respectively, are
\begin{eqnarray}
{P_3} = p^2(e) = {\left| b \right|^4}\frac{{{{\sin }^4}N\vartheta }}{{{{\sin }^4}\vartheta }}\label{same}
\end{eqnarray}
and
\begin{eqnarray}
{P_1} =\left(1- {\left| b \right|^2}\frac{{{{\sin }^2}N\vartheta }}{{{{\sin }^2}\vartheta }}\right)^2.\label{same1}
\end{eqnarray}

In the ideal case, by using a single pulse, the population inversion in the two-level system is achieved, i.e., $a=0$. This leads to $\vartheta  = \pi/2$ and $\Theta=0 $. Now let us employ the series expansion to simplify the equations in  Eqs. (\ref{alternating}), (\ref{same}), and (\ref{same1}) by ignoring the second higher order term of $a_r$ and $a_i$. Consequently, for the pulse sequence with alternating phases, we have
\begin{eqnarray}
\begin{array}{l}
{P_1} \approx 1 - 2{N^2}a_i^2,\;\;\;N = even;\\
{P_3} \approx 1 - 2{N^2}a_i^2,\;\;\;N = odd.
\end{array}
\end{eqnarray}
and for the pulse sequence with same phases, we have
\begin{eqnarray}
\begin{array}{l}
{P_1} \approx 1 - 2{N^2}a_r^2,\;\;\;\;\;\;\;\;\;\;\;\;N = even;\\
{P_3} \approx 1 - 2{N^2}a_r^2 - 2a_i^2,\;\;\;N = odd,
\end{array}
\end{eqnarray}

For both types of pulse sequences, the populations of $\left| 1 \right\rangle $ and $\left| 3 \right\rangle $ decrease as the number of pulses $N$ increases, meaning that it is easier to detect smaller static detuning when $N$ is larger. However, these two pulse sequences increase their sensitivity in different manners: the pulse sequence with alternating phases is dependent on the imaginary part of the Cayley-Klein parameter $a$, ${a_i}$, while the pulse sequence with same phases is dependent on its real part, ${a_r}$. Therefore, we can choose the proper pulse sequence to sense the effect of the static detuning deviation in terms of the size of the real and imaginary parts of Cayley-Klein parameter. In the following, we use flat $\pi$ and Gaussian pulses to examine how the effectiveness of population transfer
is affected by the static detuning deviation.

\begin{figure}[tbp]
  \centering
  \includegraphics[width=1.08\linewidth]{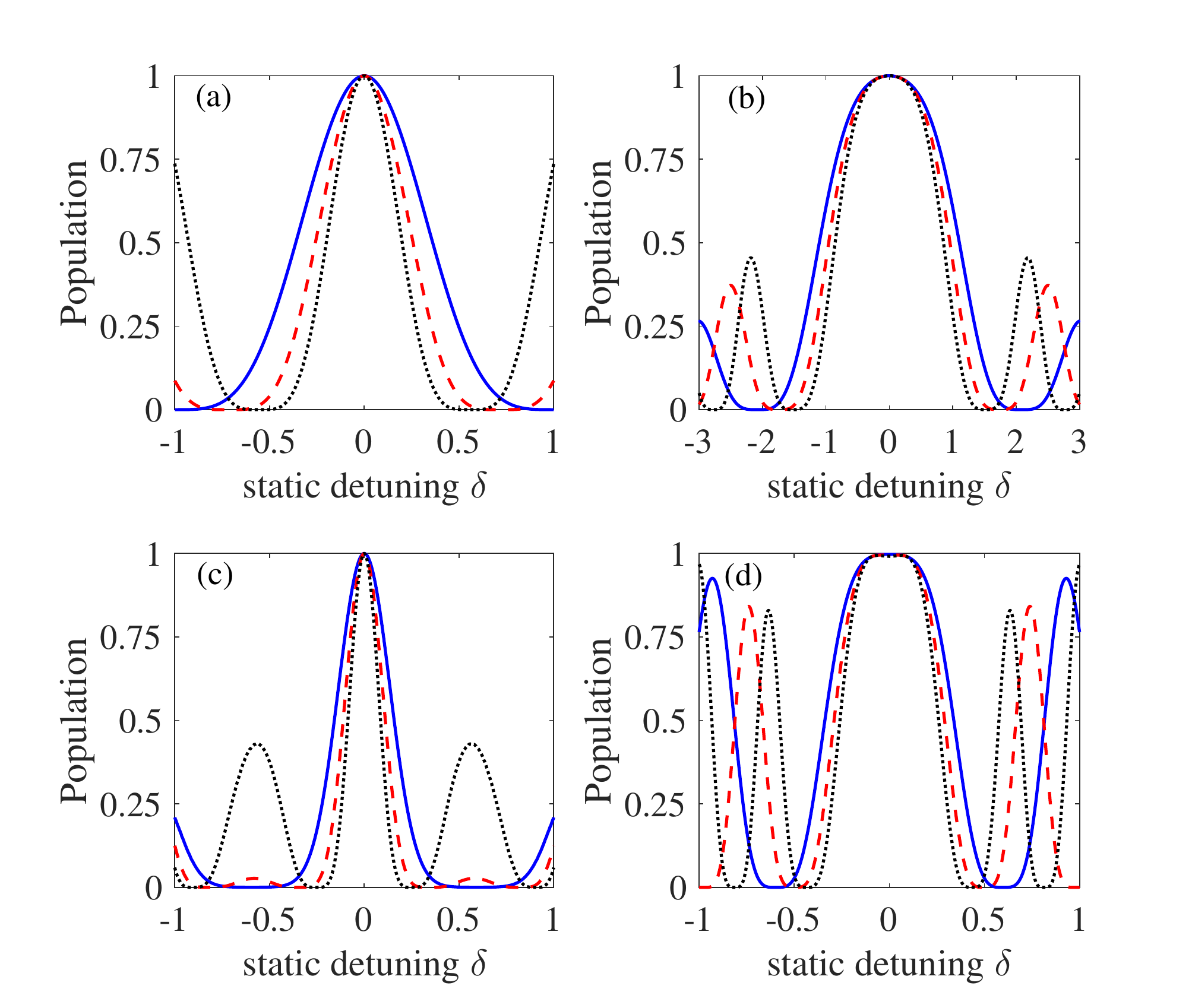}
  \caption{Population of level $\left| 3 \right\rangle $ vs the static detuning deviation $\delta$ for sequences of $N=5,$ $7,$ and $9$ identical pulses, where $N=5$ (blue, solid line), $N=7$ (red, dashed line), and $N=9$ (black, dotted line). (a) The pulse sequence with alternating phases using flat $\pi$ pulses. (b) The pulse sequence with same phases using flat $\pi$ pulses. (c) The pulse sequence with alternating phases using Gaussian pulses. (d) The pulse sequence with same phases using Gaussian pulses.}
  \label{Fig5}
\end{figure}

The first model is the flat $\pi$ pulse model, whose Hamiltonian parameters are
\begin{eqnarray}
\Delta (t) = 0,\;\;\;\Omega (t) = \sqrt 2\pi /T.
\end{eqnarray}
Its propagator has an analytical form and the corresponding Cayley-Klein parameter $a$ can be expressed as
\begin{eqnarray}
a = \cos \left( {\frac{{\sqrt {{\pi ^2} + {\delta ^2}} }}{2}} \right) + \frac{{i\delta }}{{\sqrt {{\pi ^2} + {\delta ^2}} }}\sin \left( {\frac{{\sqrt {{\pi ^2} + {\delta ^2}} }}{2}} \right).
\end{eqnarray}
Its series expansion can be written as
\begin{eqnarray}
a \approx i\frac{\delta }{\pi } - \frac{{{\delta ^2}}}{{4\pi }} + O({\delta ^3}).
\end{eqnarray}
The real and imaginary parts of $a$ are obtained as
\begin{eqnarray}
{a_r}\approx  - \frac{{{\delta ^2}}}{{4\pi }}\;,\;\;\;{a_i} \approx \frac{\delta }{\pi }\;.
\end{eqnarray}
It can be found that ${a_i}$ and ${a_r}$ is a first-order and second-order small quantities with $\delta$, respectively. This means the reduction of population caused by static detuning deviation is greater
in pulse sequence with alternating phases than the one in pulse sequence with same phases. In Figs. \ref{Fig5} (a) and (b), we plot the population $P(3)$ with the change of the static detuning deviation for $N=5$, $7$, and $9$, based on flat $\pi$ pulse sequence with alternating and same phases, respectively. One can find that under the same number of pulses $N$,
the pulse sequence alternating phases is more sensitive than the pulse sequence with same phases with respect to static detuning error.

The second model is the Gaussian pulse model. Its Hamiltonian parameters can be set as
\begin{eqnarray}
\Delta (t) = 0,\;\;\;\Omega (t) = \sqrt {2\pi }   {e^{ - {{(t/T)}^2}}} /T .
\end{eqnarray}
In Fig. \ref{Fig5} (c) and (d), we plot the population vs static detuning deviation for $N=5$, $7$, and $9$, based on Gaussian pulse sequence with alternating and same phases, respectively.
Around $\delta =0$, each of the four cases in Fig. \ref{Fig5} has a narrow feature, and it gets more narrow as the number of pulses $N$ increases. We can observe that this narrow feature is much more narrow for pulse sequences of alternating phases than that for pulses sequences of the same phases. Moreover, the feature in the Gaussian pulses near $\delta =0$ is much more narrow than the one in flat $\pi$ pulses for the same $N$. Finally, for a clearer comparison, we give the full width at half maximum of
the population change profile with respect to static detuning error near $\delta=0$ in four different schemes in Table \ref{Table2}. The full width of
half maximum of the population of flat $\pi$ pulses are more than three times the ones of Gaussian pulses for the pulse sequences of alternating phases, while they are more than twice the ones of Gaussian pulses for the pulse sequences of same phases.

\begin{table}[htb]
\centering \caption{ Full width at half maximum of population in different schemes to sense static detuning deviation. We use -A and -S to indicate the pulse sequences with alternating phases and same phases, respectively.}
\begin{tabular}{ccccc}
\hline\hline
Number of pulses     \;&  Flat $\pi$-A      \; &   Flat $\pi$-S  \;\; &  Gaussian-A    \; & Gaussian-S   \\
   \hline
$N=5$       \;&   0.72             \;&   2.21        \;\;  &  0.32         \; &     0.68      \\
$N=7$       \;&    0.52           \;&      1.91       \;\;  &      0.22      \; &      0.58    \\
$N=9$       \;&      0.40         \; &     1.71       \;\;  &      0.17     \; &   0.52        \\
\hline\hline
\end{tabular}\label{Table2}
\end{table}

\section{CASE (ii): ONE-PHOTON RESONANce}
\label{sec4}

Here, in the case of the one-photon resonance that ${\Delta _p} = {\Delta _1} = 0$, the Hamiltonian ${H_0}$ can be simplified as ${H_2}$ with
\begin{eqnarray}
{H_2} = \frac{\hbar }{2}\left( {\begin{array}{*{20}{c}}
0&{{\Omega _p}{e^{ - i\alpha }}}&0\\
{{\Omega _p}{e^{i\alpha }}}&0&{{\Omega _s}{e^{ - i\alpha }}}\\
0&{{\Omega _s}{e^{i\alpha }}}&0
\end{array}} \right),
\end{eqnarray}
In the following, we consider that the Rabi frequency error may occur on either the $P$ or $S$ pulse alone. If the $S$ or $P$ pulse has an error, i.e.,
\begin{eqnarray}
{{\Omega _i}} \to (1 + \eta ){{\Omega _i}}.
\end{eqnarray}
where $i=s,p$. We will call ${{\Omega _i}}$ the standard pulse and ${{\Omega' _i=(1 + \eta ){{\Omega _i}}}}$ the error pulse.
Firstly, when ${{\Omega _p}}$ and ${{\Omega _s}}$ are constants, the propagator of the system can be expressed as
\begin{eqnarray}
{U_2} = \left( {\begin{array}{*{20}{c}}
{{{[{U_2}]}_{11}}}&{{{[{U_2}]}_{12}}}&{{{[{U_2}]}_{13}}}\\
{{{[{U_2}]}_{21}}}&{{{[{U_2}]}_{22}}}&{{{[{U_2}]}_{23}}}\\
{{{[{U_2}]}_{31}}}&{{{[{U_2}]}_{32}}}&{{{[{U_2}]}_{33}}}
\end{array}} \right),
\end{eqnarray}
where
\begin{eqnarray}
\begin{array}{l}\vspace{0.06cm}
{{{[{U_2}]}_{11}}} = (\Omega _s^2 + \Omega _p^2\cos A)T^2/{A^2},\\\vspace{0.06cm}
{{{[{U_2}]}_{22}}} = \cos A,\\\vspace{0.06cm}
{{{[{U_2}]}_{33}}}= (\Omega _p^2 + \Omega _s^2\cos A)T^2/{A^2},\\\vspace{0.06cm}
{{{[{U_2}]}_{12}}}={{{[{U_2}]}_{21}}}={{{[{U_2}]}_{23}}}={{{[{U_2}]}_{32}}} =  - i{\Omega _p}T\sin A/A,\\\vspace{0.06cm}
{{{[{U_2}]}_{13}}} ={{{[{U_2}]}_{31}}}= {\Omega _p}{\Omega _s}(\cos  A - 1)T^2/A^2.
\end{array}
\end{eqnarray}
Here $A = \sqrt {{{{\Omega _p}}^2T^2} + {{{\Omega _s}}^2T^2}}/2 $. It can be found that the solution to achieve complete population transfer without error is
\begin{eqnarray}
{{\Omega _p}} = {{\Omega _s}} = \sqrt 2 \pi /T,
\end{eqnarray}
where $T$ is the pulse width.
\begin{figure}[tbp]
  \centering
  \includegraphics[width=1.08\linewidth]{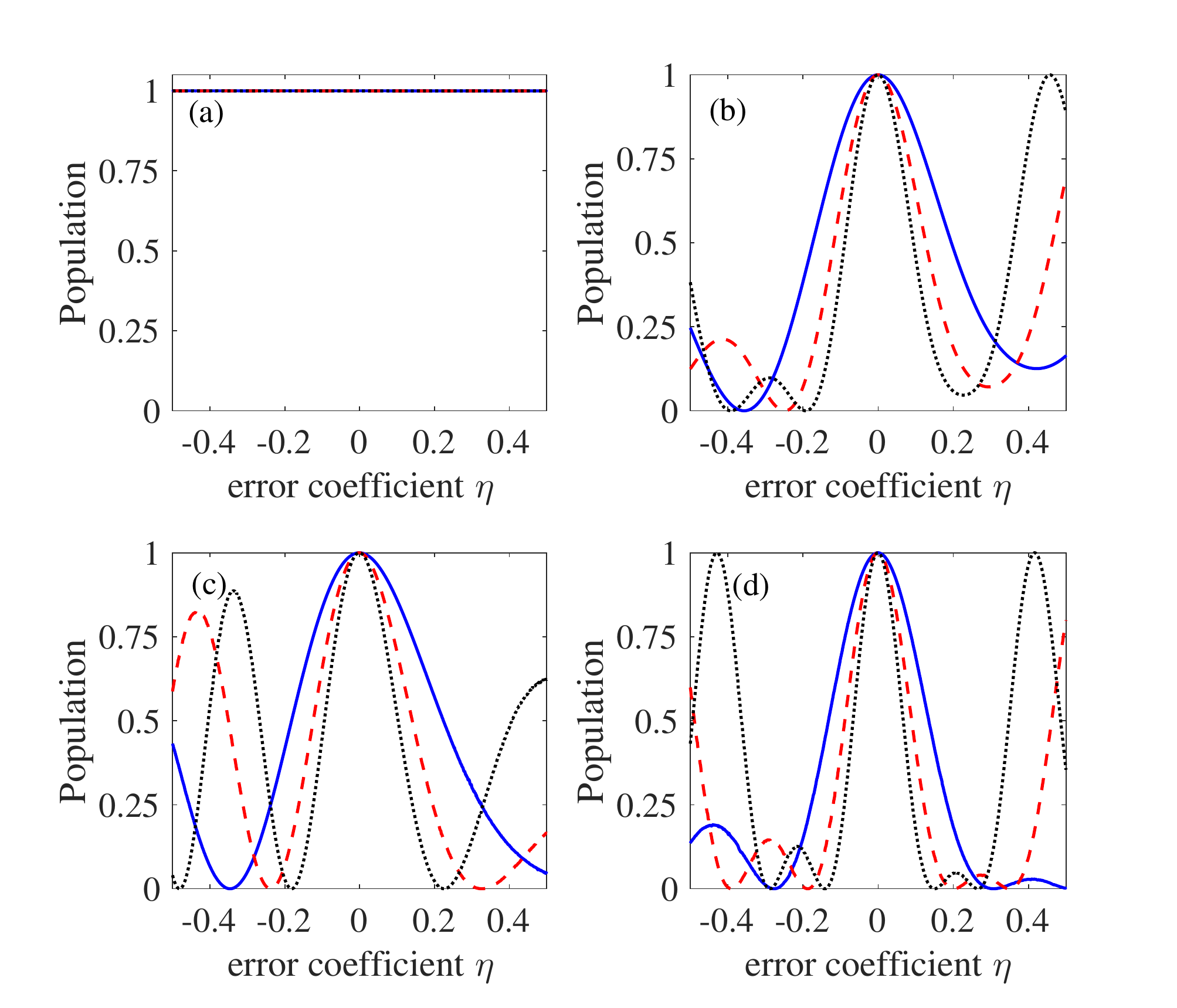}
  \caption{Population of level $\left| 1 \right\rangle $ vs the Rabi frequency error $\eta$ for sequences of $N=4,$ $6,$ and $8$ identical pulses, based on CDS, where $N=4$ (blue, solid line), $N=6$ (red, dashed line), and $N=8$ (black, dotted line). (a) The pulse sequence with alternating phases for fixed error. (b) The pulse sequence with same phases for fixed error. (c) The pulse sequence with alternating phases for alternating error. (d) The pulse sequence with same phases for alternating error.}
  \label{Fig6}
\end{figure}
 We denote this scheme as the constant driven scheme (CDS). It is worth pointing out that the error pulse can be applied to one of $P$ and $S$ all the time (fixed error case), or applied alternatively to the pulses between $P$ and $S$ (alternating error case). Specifically, the alternating error uses the error pulse $\Omega' _s$ in the first pulse, $\Omega' _p$ in the second pulse, $\Omega' _s$ in the third pulse, and so on.

Here, we study how the population of the level
$\left| 1 \right\rangle $ are affected by the Rabi frequency error for even-numbered pulses, $N=4$, $6$, and $8$ in CDS for four cases: (i) The pulse sequence with alternating phases for fixed error;
(ii) The pulse sequence with same phases for fixed error; (iii) The pulse sequence with alternating phases for alternating error;  (iiii) The pulse sequence with same phases for alternating error.
For the case (i), we can find that
\begin{eqnarray}
{U_2}(\eta ,0){U_{2, - }}(\eta ,0) = {U_2}(0,\eta ){U_{2, - }}(0,\eta ) = I,
\end{eqnarray}
where ${U_2}(\eta,0)$ and ${U_{2, - }}(\eta,0)$ represent the propagator that there is a Rabi frequency error in a $P$ pulse with zero and $\pi$ phases, respectively.
${U_2}(0,\eta)$ and ${U_{2, - }}(0,\eta)$ represent the propagator that there is a Rabi frequency error in a $S$ pulse with zero and $\pi$ phases, respectively. This implies that
the sensitivity to Rabi frequency error will not increase with the increase of the number of pulses in the case the pulse sequence with alternating phases for fixed error, shown in Fig. \ref{Fig6} (a).
 In Fig. \ref{Fig6} (b-d), we plot the performance of population with respect to $\eta$ for other three cases. We can see that the sensitivity increases with the increase of the number of pulses in the case (ii), (iii), and (iiii). Among them, the  pulse sequence with same phases for alternating error shows the highest sensitivity.

As a comparison, we use the stimulated Raman adiabatic passage (STIRAP) \cite{STIRAP1,STIRAP2,super} to investigate the population change with respect to the Rabi frequency error.
In traditional STIRAP, $P$ pulse and $S$ pulse are usually identical symmetric
functions of time but implemented in counterintuitive order, i.e.,
\begin{eqnarray}
{{\Omega _p}} = {\Omega _0}f(t),\;\;\;{{\Omega _s}} = {\Omega _0}f(t - \tau ),
\end{eqnarray}
where ${\Omega _0}$ and $\tau$ are the peak Rabi frequency and pulse delay, respectively. $S$ pulse is implemented earlier than $P$ pulse. If a pair of $P$ and $S$ pulses can achieve a complete population transfer of $\left| 1 \right\rangle  \to \left| 3 \right\rangle $, then one can change the time order of $P$ and $S$ pulses to achieve the transfer of $\left| 3 \right\rangle $ back to $\left| 1 \right\rangle $.
However, adiabatic protocols usually require higher pulse peak to achieve high fidelity population transfer. If we adopt the adiabatic pulse schemes with lower pulse peak, we usually cannot achieve the above population transfer process.
Here, adopting the adiabatic pulse method with low peak value, we consider the pulse sequence with alternating phases. The reasons is as follows.
Without Rabi frequency error, we can drive the system completely to return to the initial state $\left| 1 \right\rangle$ after executing a pair of STIRAP with opposite phases. It usually can not completely return to $\left| 1 \right\rangle$ in the presence of this error. This means that we can detect this error by measuring the population change of $\left| 1 \right\rangle$.
Also, the case of fixed error is not considered here due to the fact that
\begin{eqnarray}
{U_2}(\eta,0){{\tilde U}_{2, - }}(\eta,0) = {U_2}(0,\eta ){{\tilde U}_{2, - }}(0,\eta ) = I.
\end{eqnarray}
where the tilde indicates the propagator that the time order of $P$ and $S$ is exchanged.
For the STIRAP, we make use of the pulse sequence with alternating phases for alternating error to sense Rabi frequency error by applying several common pulses: Gaussian pulses, hyperbolic secant
(sech) pulses, and sinusoidal pulses.

\begin{figure}[tbp]
  \centering
  \includegraphics[width=1.08\linewidth]{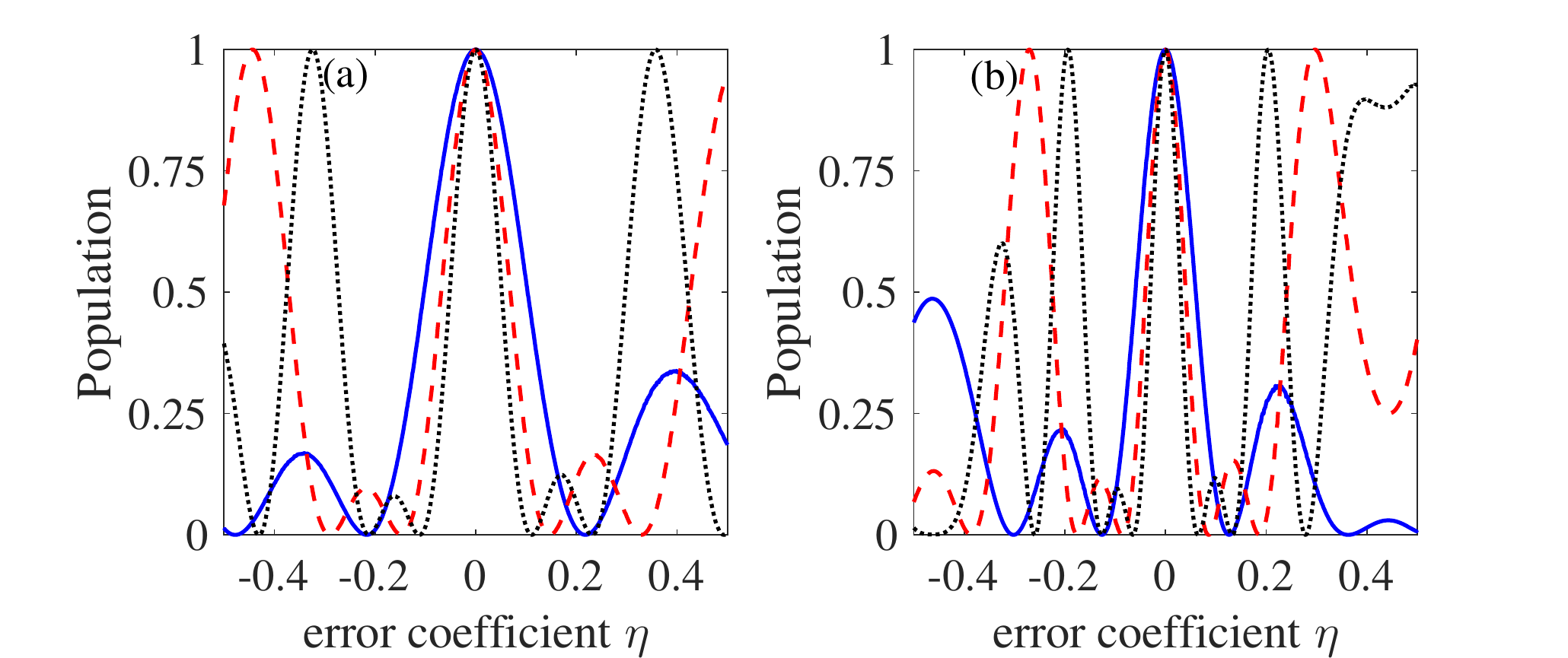}
  \caption{Population of level $\left| 1 \right\rangle $ vs the Rabi frequency error coefficient $\eta$ for sequences of $N=4,$ $6,$ and $8$ identical pulses, where $N=4$ (blue, solid line), $N=6$ (red, dashed line), and $N=8$ (black, dotted line). (a) The pulse sequence with alternating phases for alternating error using Gaussian pulses. (b) The pulse sequence with alternating phases for alternating error using $\rm {sech}$ pulses.}
  \label{Fig7}
\end{figure}

\begin{table*}[htb]
\centering \caption{ Full width at half maximum of population in different
schemes to sense the Rabi frequency error. We use -SF, -AA, and -SA to indicate the pulse sequence with same phases for fixed error, alternating phases for alternating error, and same phases for alternating error, respectively.}
\begin{tabular}{cccccccc}
\hline\hline
Number of pulses   \; \;\;\;\;\; & CDS-SF             \; \;\;\;\;\;&  CDS-AA        \; \;\;\;\;\; & CDS-SA     \;\;\;\;\;\;  &    Gaussian-AA   \;\;\;\;\;\; &  sech-AA   \; \;\;\;\;\;&  sin-AA    \;\;\;\;\;\; &  $\sin^2$-AA \\
   \hline
$N=4$         \;\;\;\;\;&  0.369             \;\;\;\;\;&  0.406            \;\;\;\;\;&  0.267          \;\;\;\;\;&         0.206   \;\;\;\;\;&    0.120    \;\;\;\;\;&  0.561       \;\;\;\;\;&  0.720       \\
$N=6$          \;\;\;\;\;&  0.225             \;\;\;\;\;&  0.264            \;\;\;\;\;&  0.178          \;\;\;\;\;&          0.137 \;\;\;\;\;&     0.080    \;\;\;\;\;&  0.375       \;\;\;\;\;&  0.482       \\
$N=8 $        \;\;\;\;\;&  0.184             \;\;\;\;\;&  0.198            \;\;\;\;\;&  0.184          \;\;\;\;\;&         0.103\;\;\;\;\;&        0.060   \;\;\;\;\;&  0.282       \;\;\;\;\;&  0.362       \\
\hline\hline
\end{tabular}\label{Table3}
\end{table*}

The Rabi frequencies of Gaussian pulses are
\begin{eqnarray}
{{\Omega _p}} = {\Omega _0}{e^{ - {{(t/T)}^2}}},\;\;\;{{\Omega _s}} = {\Omega _0}{e^{ - {{(t/T - 2)}^2}}},
\end{eqnarray}
and the Rabi frequencies of hyperbolic secant pulses are
\begin{eqnarray}
{{\Omega _p}} = {\Omega _0}{\mathop{\rm sech}\nolimits} (t/T),\;\;\;{{\Omega _s}} = {\Omega _0}{\mathop{\rm sech}\nolimits} (t/T - 5),
\end{eqnarray}
respectively. Here, the amplitude of Rabi frequency is set as ${\Omega _0}=\sqrt 2 \pi /T$.
In Fig. \ref{Fig7}, we plot the population of the level $\left| 1 \right\rangle$ with respect to the Rabi frequency error for $N = 4$, $6$, and $8$ pulses in
pulse sequence with alternating phases for alternating error using Gaussian pulses and $\rm {sech}$ pulses, respectively. Their population profiles are influenced largely by
the Rabi frequency error. We can see that each of them forms a narrow feature near $\eta=0$, which gets more narrow with the increase of the number of pulses $N$.
Furthermore, the feature of the sech pulse sequences shows more narrow than that of Gaussian pulse sequences near $\eta=0$.

\begin{figure}[tbp]
  \centering
  \includegraphics[width=1.08\linewidth]{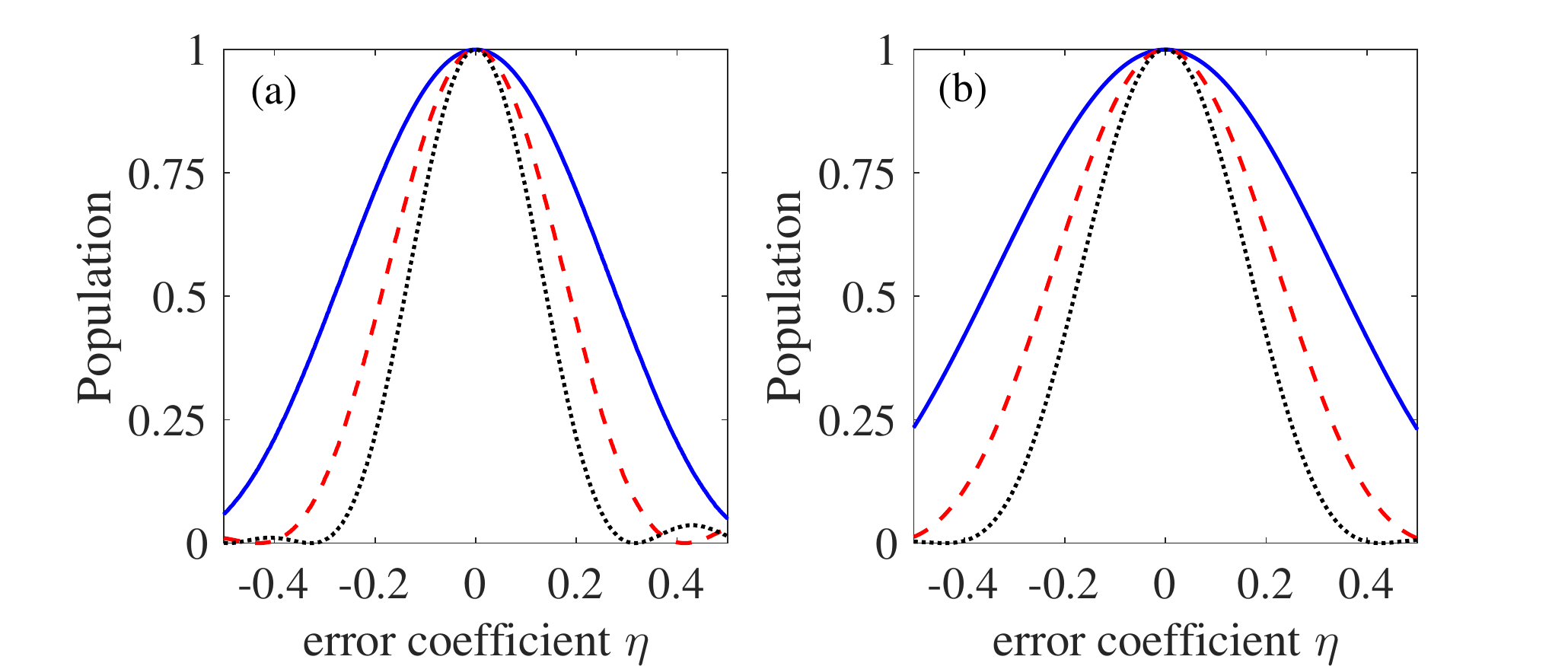}
  \caption{Population of level $\left| 1 \right\rangle $ vs the Rabi frequency error coefficient $\eta$ for sequences of $N=4,$ $6,$ and $8$ identical pulses, where $N=4$ (blue, solid line), $N=6$ (red, dashed line), and $N=8$ (black, dotted line). (a) The pulse sequence with alternating phases for alternating error using $\sin $ pulses. (b) The pulse sequence with alternating phases for alternating error using ${\sin ^2}$ pulses.} 
  \label{Fig8}
\end{figure}

Finally, we consider two types of sinusoidal pulses for comparison. The Rabi frequencies of the first sinusoidal pulses are
\begin{eqnarray}
{{\Omega _p}} = {\Omega _0}\sin (\pi t/T),\;\;\;{{\Omega _s}} = {\Omega _0}\cos (\pi t/T).
\end{eqnarray}
The Rabi frequencies of second sinusoidal pulses take the form of
\begin{eqnarray}
{{\Omega _p}} = {\Omega _0}{\sin ^2}(\pi t/T),\;\;\;{{\Omega _s}} = {\Omega _0}{\cos ^2}(\pi t/T),
\end{eqnarray}
where ${\Omega _0}=\sqrt 2 \pi /T $. In Fig. \ref{Fig8}, we plot the population of the level $\left| 1 \right\rangle$ with respect to the Rabi frequency error for $N = 4$, $6$, and $8$ pulses in
pulse sequence with alternating phases for alternating error using these two sinusoidal pulses, respectively.
It can be seen that the sensitivity with Rabi frequency error increases with the increase of the number of pulses, and the sensitivity of the sin pulse sequence is higher than that of $\sin^2$ pulse sequence.

At the end of this section, we summarize the performances of these different schemes by comparing the full width at half maximum of the population change curve with respect to Rabi frequency error near $\eta=0$ when the number of pulses is same, shown in Table \ref{Table3}. Among them, sech pulse sequence with alternating phases for alternating error shows the highest sensitivity, followed by the Gaussian pulses and then the CDS protocol. The sinusoidal pulses methods take the highest values of the full width at half maximum of the population change curve.

\section{CONCLUSION}
\label{sec5}
In conclusion, we study the sensitivity of composite pulse sequences to parameter fluctuations in two families of Hamiltonian with the SU(2) dynamic symmetry: off-resonance case and one-photon resonance case, using some popular coherent control techniques. All of control pulses can realize high-fidelity population transfer in three-level quantum systems. The repeated application of the same
gate-generating pulse greatly amplifies the effect of the experimental errors originating from the imperfection of parameter control and the error effect is mapped
onto the change of populations. It is found that the sensitivity with respect to these errors are greatly improved with the increase of the number of pulses. In the first case, the pulse sequence with same phases is more sensitive to Rabi frequency error than that with alternating ones, and $\pi$ pulse sequence is the most sensitive among the considered coherent control schemes. For static detuning deviation, we find that the pulse sequence with alternating phases is more sensitive than that with same ones, and the Gaussian pulse sequence is more sensitive than the flat $\pi$ pulse sequence.
In the second case, we use the pulse sequence with alternating phases to study the sensitivity with respect to Rabi frequency error. By comparing the full width at half
maximum of the population change in the different control schemes, we find that the sech pulse sequence shows the highest sensitivity, followed by the Gaussian pulses, then
CDS protocol, and finally the sinusoidal pulses.

There is still much room for improvement in the work of using composite pulse sequence to realize quantum sensing. One can design an optimal composite pulse sequence that is highly sensitive to the specific parameter in physical systems and can suppress the influence of other parameters at the same time. Moreover, only two kinds of composite pulse sequences are considered in this paper, and there may still be more complex composite pulse sequences that are more sensitive to experimental errors. Finally, the composite pulse schemes in different quantum systems show different performances, thus one can design a simple and suitable pulse sequence to improve the error sensitivity in terms of the properties of these systems.

\section*{ACKNOWLEDGMENTS}
This work is supported by the National Natural Science Foundation of China (Grants No. 12004006, No. 12075001, and No. 12175001), and Anhui Provincial Natural Science Foundation (Grant No. 2008085QA43).

\end{document}